# "Nice to meet you!": Expressing Emotions with Movement Gestures and Textual Content in Automatic Handwriting Robots

Yanheng Li
City University of Hong Kong
Hong Kong, Hong Kong
Lydia.YH-Li@my.cityu.edu.hk

Lin Luoying
City University of Hong Kong
Hong Kong, Hong Kong
Lin.luoying@my.cityu.edu.hk

Xinyan Li
City University of Hong Kong
Hong Kong, Hong Kong
xinyali9-c@my.cityu.edu.hk

Yaxuan Mao
City University of Hong Kong
Hong Kong, Hong Kong
yaxuanmao2-c@my.cityu.edu.hk

RAY LC
City University of Hong Kong
Hong Kong, Hong Kong
LC@raylc.org

## ABSTRACT

Text-writing robots have been used in assistive writing and drawing applications. However, robots do not convey emotional tones in the writing process due to the lack of behaviors humans typically adopt. To examine how people interpret designed robotic expressions of emotion through both movements and textual output, we used a pen-plotting robot to generate texts by performing human-like behaviors like stop-and-go, speed, and pressure variation. We examined how people convey emotion in the writing process by observing how they wrote in different emotional contexts. We then mapped these human expressions during writing to the handwriting robot and measured how well other participants understood the robot's affective expression. We found that textual output was the strongest determinant of participants' ability to perceive the robot's emotions, whereas parameters of gestural movements of the robots like speed, fluency, pressure, size, and acceleration could be useful for understanding the context of the writing expression.

## CCS CONCEPTS

• **Computer systems organization** → Robotics; • **Human-centered computing** → User studies.

## KEYWORDS

handwriting robot, robot emotional expression, robotic gestures

## 1 INTRODUCTION

Handwriting robots have been used to produce automatic writing for applications like dictation, copying, artistic expression, and error-free operation.

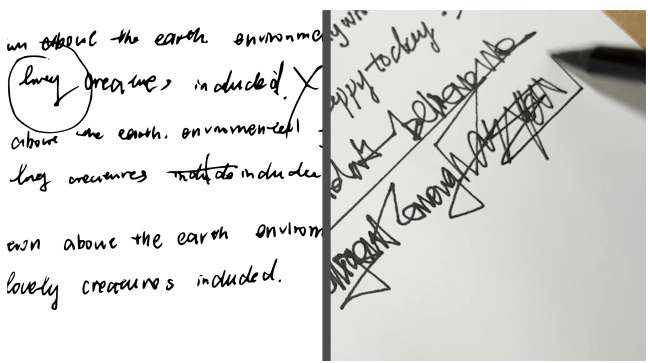

Figure 1: (Left) A participant writes in anger through the emotional expression of fast speed, aggressive symbols, and strike-through. (Right) The axidraw robot attempts to express anger by mimicking emotional expressions through writing movements.

Research on calligraphy robots has explored techniques for robot motion planning and creation by learning and imitating the trajectory of human writing through machine learning [20]. Writing robots have also been applied to developing children's handwriting skills by engaging them in social interactions and robot movements [6]. Emotional expression through handwriting robots can potentially help people with disabilities express themselves effectively [10, 21]. Artists have also employed handwriting robots to carry out co-creation work and reflect on the role of human beings in the process [11, 16, 18].

Beyond only having handwriting robots produce content to assist people's self-expression, the writing performance of robots itself has also begun to be examined as a way of expressing emotion. Robots may reveal a particular emotional state to people by changing different gestures and movement patterns, thus influencing people's emotional responses and reactions [1, 14, 15]. These emotional gestural movements can be used in turn to vividly and accurately represent people's thoughts and affection. When people communicate in text, they try to develop various ways to compensate for the lack of nonverbal features in verbal expressions, such as making use of punctuation and emoticons, in addition to the text [9]. We study whether robots can effectively leverage these types of nuanced emotional expressions through gestural movements and textual information.

Different emotional handwriting features, such as baseline, slant, pressure, size, spacing, zone, speed, pause, and connecting, are able to reflect a person's emotional states and personality [7, 13]. Since handwriting robots are equipped with motor-driven arms and "hand" holding pens, motor movements of robot writing systems can potentially offer the same emotional features [19].

Databases in previous research on human writing have only image data [13]. Further exploration is needed to show whether human emotions can be effectively perceived through the representation of robot writing. Our research questions:

- How can we design the movement of the writing robot to express emotions? What kind of movement parameters can best demonstrate the robot's emotions during movement?
- What textual information can best clarify the emotion of the robot?
- Do gestural movements and textual information have any intersecting effect on human interpretation of robot emotions?

First, we invited participants in a pre-design to write passages on paper of specified emotional conditions and explain the emotional features of their handwriting. Then we abstracted the observed writing patterns, combining them with robot emotional gestures for the writing robot to imitate. Next, we invited a different group of participants to evaluate the effectiveness of emotion conveyance and how they recognized different emotional writing. Understanding how handwritten text and gestural movements can communicate emotional information allows designers to apply these communication strategies to indicate emotional content from text production alone, which would benefit assistive writing applications.

## 2 METHODS

We modeled four emotions based on the circumplex model of emotion [17]: happiness, anger, sadness, and calmness. These four emotions were selected due to that they are four different types of emotion in a two-dimensional circular space. Happiness has a positive valence and high arousal; anger has a negative valence and high arousal; sadness has a negative valence and low arousal, while calmness has a positive valence and low arousal. We also set a neutral emotion condition as a control set of parameters.

### 2.1 Pre-design

To capture the fine motor dexterity of humans for different emotional expressions in the form of video, we invited 6 participants to write certain sentences in the patterns of the four emotions we pre-selected. Participants were given a sentence without emotion explaining a concept of science. Before writing with emotional patterns, they were asked to write a few times in case unfamiliarity affected the emotional representations. Compared with the features mentioned in previous research about the emotional gestural movements of robots and affective human writing, the accessible writing movement patterns of the participants in our pre-design are as followed: velocity (high arousal: fast, low arousal: slow), acceleration (the speed is stable in low valence and low arousal) [12], force (high valence: slight, low valence: heavy; high arousal: heavy, low arousal: slight), fluency (write in smooth connection, sadness is not fluent while happiness, for example, is fluent), size (high arousal: big, low arousal: small) [8, 13]; the textual information includes punctuation, emoticons (e.g., "!" shows happiness, "..." means thinking in calm or sadness and silence) [2, 4] and the content of the verbal text.

Table 1: Robot movement design examples and parameters.

| Emotion | Writing Sample | Emotional Pattern |
|---|---|---|
| Neutral Line 1 | | Speed: 25 (1-100), Accel: 75 (1-100) Force: 70 (1-100), Size: small Fluency: not so fluent |
| Happy Line 2 | | Speed: 75, Accel: 85 Force: 50, Size: medium Fluency: connect fluently |
| Happy Line 4 | | Speed:100, Accel: 100 Force:80, Size: big Fluency: all connect together |
| Happy Line 6 | | Speed: 75, Accel: 85 Force:50, Size: medium Fluency: connect fluently |
| Sad Line 3 | | Speed: 15, Accel: 0 Force: 60 Fluency: connect abruptly |
| Angry Line 5 | | Speed: 100, Accel: 100 Force: 80, Size:big Fluency: all connect together |
| Calm Line 7 | | Speed: 25, Accel: 60 Force: 70, Size: small Fluency: more fluent than default |

### 2.2 Movement gesture design

As shown in Table 1, we applied the gestural movements with different parameters for each sentence of the four emotions, using Axidraw V3, a pen-and-paper X-Y plotter to present [3, 5]. For neutral emotion, we keep all the default movement parameters to control the Axidraw. Based on the pre-design evaluation, we set the parameters as follows. For happiness, we make faster speed and higher acceleration, and lift the pen to lower the pressure; we also set the robot with fluent movement and rounded connection so that the robot can write in a brisk movement conveying an eased mood. For anger, we maximize the speed and acceleration and keep the pen position on the paper to simulate an impatient, messy, and uninterrupted movement. For sadness, we make the speed uniform and as slow as possible. The robot writes casually and abruptly with heavier pressure than in the happiness condition to show its depression and unwillingness to do anything. For calmness, we try to create a character that the robot is taking the writing seriously and concentrating on the task, so the speed is not so fast and works more fluently than the default.

Table 2: The design of textual information.

| Emotion | Writing Sample | Emotional Pattern |
|---|---|---|
| Neutral Line 1 | | Text: typography |
| Happy Line 2 | | Text: handwriting, happy expression Punctuation: "!" shows happy |
| Happy Line 4 | | Text: handwriting, happy emotional word, Emoticons: ": )" |
| Happy Line 6 | | Text: handwriting, happy expression Punctuation: tilde |
| Sad Line 3 | | Text: handwriting, sad expression Punctuation: "..." shows depressio |
| Angry Line 5 | | Text: careless handwriting, chaotic erase |
| Calm Line 7 | | text: handwriting, emotional word "calm", clear erase, Punctuation: "..." shows calm |

### 2.3 Textual information design

To explore how users distinguish emotions from the different textual content and how verbal and nonverbal emotional features may intersect to influence people's comprehension, we also divided the text into some including emotional words, and others without emotional words as shown in Table 2.

### 2.4 User Study procedure

The evaluation study was set up as an online survey based on videos of the actual performance of the robot in action. We invited 17 college students including 8 females and 9 males from 21 to 29 years old ($M = 23.88$, $SD = 2.42$) to evaluate what they think the robot feels in writing and which designed factor influences their perception of the robot's affection the most. Over half (58.82%) have interacted with robots in general, while 17.65% of participants have no experience working with robots. 5.88% have some engineering experience. Participants were directed to a website that gave instructions for the study. After giving consent, participants started the survey. The total time for completion was twenty minutes.

We used a within-subjects repeated measures design; all the participants watched all the writing samples of the four emotions. The same set of questions had to be completed for each video of the different robot movement expressions. We carried out an evaluation of (1) whether the subjects could accurately recognize the emotions expressed by the Axidraw based on the gestural movement parameters and the textual output, and (2) how influential each parameter in both movements and textual information was in the emotional expression of the robot.

After analyzing quantitative data obtained from the online survey, we also conducted an informal discussion with participants to reveal details of expressivity in the robot, and the intersecting effects of text and gestural movements in the third research question.

## 3 RESULTS

For the general perception of robot writing emotions, it appears that participants are able to perceive emotions well, as shown in Fig. 2. Among all the seven lines, only line 1 appeared to be misunderstood as "calm" by the majority of the subjects (58.82 ± 3.68%, 95% CI), with "neutral" being the correct answer recognized by 35.29% of all, while the average correct recognition rate of the real calm line 7 is around 58.82%. The most easily accessible emotion is "sad," with 70.59% of participants correctly decoding. Lines 2, 4, and 6 all present the "happy" emotion with the average proportion of accurate recognition being around 60.78%, and the rate of "angry" is 64.71%. Overall, the data shows the difficulties in identifying calm emotions while extreme emotions, like sadness or anger, are more obvious and were perceived correctly more often by participants.

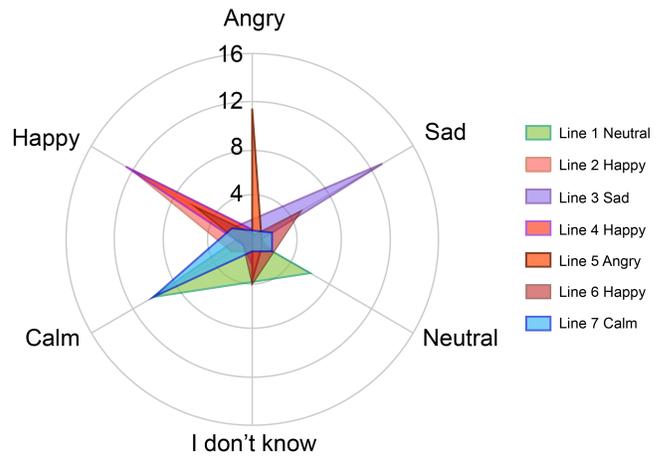

Figure 2: Line 1 is designed to express neutral emotion, but the green zone shows that more participants believed it is a calm emotion. Lines 2, 4, and 6 should be the happy expressions; the red zones indicate that the participants can feel the emotion we intended. Line 3 should be sadness; the participants' response is also in line with our intended expression, as seen in the purple zone. For angry line 5, over 10 participants think it is angry according to the orange zone. For calm line 7, it is shown in the blue zone how well they can recognize the calm emotional expression.

To determine which handwriting parameters have an impact on the participant's perception of the robot's emotions, we used an indirect self-report approach due to the limitations of an online study. We adapted 7-point Likert scales to probe whether the participants agree or disagree with the statement that particular parameters influence their perception of emotions.

Fig. 3 shows that the textual factors exhibited an overwhelming advantage in passing on emotions. Emotional words have a strong impact on human perception, followed by the content of texts without emotional words, emoticons, and punctuation.

In terms of gestural movements, the most influential parameter is speed (71% agreed), following are fluency (65%), pressure (59%), size (59%) and acceleration (53%).

One-way ANOVA reveals statistically significant differences amongst the 9 factors listed above (p = 0.0038), suggesting that there is a significant difference amongst the different parameters in subjects' perceived ability to distinguish emotion based on these parameters, i.e. the parameters are not all the same for subjects.

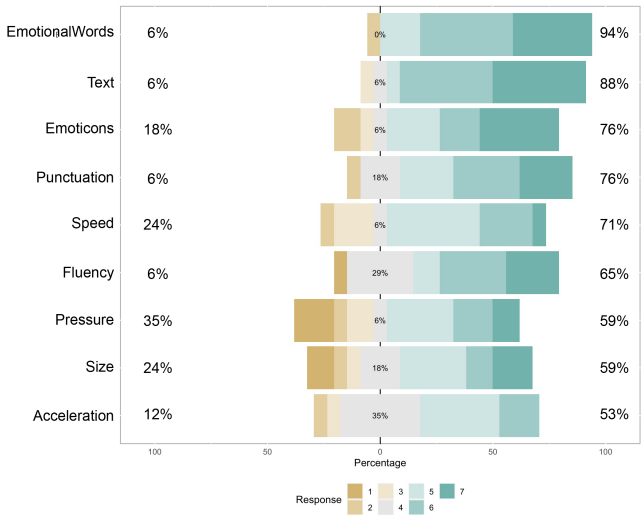

**Figure 3: 1 point means the participants strongly disagree with the referred parameter having an effect on their perception of emotional expression in the video, while 7 points means strongly agree. For example, we asked the participants about particular parameters like "You can realize the robot is expressing [a particular emotion] through its moving speed," and asked participants to rate their agreement with this particular assertion.**

From the short answer questions in the survey and informal discussion afterward with participants, we noticed that people could recognize specific movement parameters (e.g., fast speed) and unique punctuation representing a specific arousal or valence level more precisely, rather than decide on a particular emotion. However, subtle expressions may be conveyed through the outcome textual output. We observed that the exclamation mark "!" shows happiness or anger, and the more exclamation marks, the higher the arousal level of the emotion is. Meanwhile, we also found that the ellipsis means sadness or anger. Regarding speed, a majority(>10) of participants sensed anger or happiness when the robot wrote quickly and discerned sadness when it wrote slowly. There were also participants who suggested that speed can only indicate the arousal level of the emotion but not the valence of the emotion. One participant said, "fast speed indicates a calm emotion, while slow speed indicates excited emotion. I cannot tell whether it is happy." Most participants stated that they do not think acceleration makes a big difference; only one mentioned that the variation of the speed in the writing process could indicate the pleasure level of the emotion. When the acceleration is high, the emotion appears to be interpreted negatively, such as anger.

Interestingly, four participants mentioned that they believe the textual information comes prior to the emotional perception, enhancing human comprehension of emotional movement presented later. One participant said she recognized the happiness of line 6 through the tilde and could clarify the emotion of the movement.

## 4 DISCUSSION, LIMITATION, FUTURE WORK

Emotional expression in handwriting robots allows textual communication with stylized content, leading to the application of assistive technologies for affective communication. We find that textual information like emotional words and texts appear to serve an informative role in showing the robot's emotions. Preliminary work shows that emoticons and punctuation provide help in enhancing the perception of arousal or valence. On the other hand, speed, fluency, pressure, and size effectively influence human perception of robot emotion, while acceleration displays less influence. Speed is not only useful for detecting emotion but also contributes to the valence of the response (for example, happiness vs anger).

The screen-based evaluation exhibits some limitations. The participants may need help to assess the actual interaction with the system. The parameters we measured with the participants are all chosen by us in advance. In contrast, the in-person evaluation would allow the participants to self-report another parameter we did not cover in the research. Moreover, we carried out a within-subject experiment by designing three happy lines with different patterns due to the many parameters and conditions for each emotion that needed to be explored exhaustively. Although there are three sentences indicating happiness, their content is different. Since the evaluation of each subsequent emotionally expressive movement may affect further emotional evaluations in a carry-over effect, future work should also control the setting of each parameter in a between-subject experiment for a better-controlled evaluation.

One goal of this study is to explore effective emotional writing movements presented by handwriting robots. However, the results show that textual information is the determining factor for expressing emotions. The importance of other handwriting features needs to be further studied. One of the unanticipated parameters is the sounds made by the writing machine during the working process. According to one participant, sounds play a more significant part in influencing his evaluation of emotion identification than typography. For example, when the robot moves at a certain speed or across a certain range, the volume and frequency of noises would provide people with the mood of the writing machine, especially negative emotions. Future studies need to examine robots as holistic machines that use visual, auditory, and spatial cues to affect emotional perception, just as situated human peers affect our perception in multidimensional settings.

In the future, we hope to examine how the emotions of writing robots can assist people with collaborative text generation with emotional features. In addition, we can broaden the type of texts we are asking robots to generate to allow a more interactive writing format. For example, we can create a Machine Learning system to help the robots generate texts for emotional expression.